\documentclass[a4paper,twoside]{article}

\usepackage{epsfig}
\usepackage{subcaption}
\usepackage{calc}
\usepackage{amssymb}
\usepackage{amstext}
\usepackage{amsmath}
\usepackage{amsthm}
\usepackage{booktabs}
\usepackage{multicol}
\usepackage{pslatex}
\usepackage{apalike}
\usepackage{algorithm2e}
\usepackage{tikz}
\usepackage[bottom]{footmisc}
\usepackage{SCITEPRESS}     
\usepackage[justification=centering]{caption}
\usepackage{float}
\usepackage{hyperref}
\usepackage{balance}

\begin{document}

\title{Auto-grader Feedback Utilization and Its Impacts: \linebreak An Observational Study Across Five Community Colleges}

\author{\authorname{Adam Zhang\orcidAuthor{0009-0004-3791-7311}, Heather Burte\orcidAuthor{0000-0002-9623-4375}, Jaromir Savelka\orcidAuthor{0000-0002-3674-5456}, Christopher Bogart\orcidAuthor{0000-0001-8581-115X}, and  Majd Sakr\orcidAuthor{0000-0001-5150-8259}}
\affiliation{School of Computer Science, Carnegie Mellon University, Pittsburgh, Pennsylvania, U.S.A.}
}


\abstract{Automated grading systems, or auto-graders, have become ubiquitous in programming education, and the way they generate feedback has become increasingly automated as well. However, there is insufficient evidence regarding auto-grader feedback's effectiveness in improving student learning outcomes, in a way that differentiates students who utilized the feedback and students who did not. In this study, we fill this critical gap. Specifically, we analyze students' interactions with auto-graders in an introductory Python programming course, offered at five community colleges in the United States. Our results show that students checking the feedback more frequently tend to get higher scores from their programming assignments overall. Our results also show that a submission that follows a student checking the feedback tends to receive a higher score than a submission that follows a student ignoring the feedback. Our results provide evidence on auto-grader feedback's effectiveness, encourage their increased utilization, and call for future work to continue their evaluation in this age of automation}

\onecolumn \maketitle \normalsize \setcounter{footnote}{0} \vfill

\section{\uppercase{Introduction}}
\label{sec:introduction}
Automated grading systems (auto-graders), are a popular solution to providing immediate scores and feedback in programming courses. Though auto-graders may not consistently provide the same quality of feedback as a human instructor, they can still be valuable to instructors and students alike due to their immediacy and constant availability. Auto-graders' scalability supports teaching larger classes, and their efficiency provides students with more timely, on-going guidance for learning and problem-solving. However, empirical assessments of auto-grader feedback's impact on learning outcomes, such as grades and pass rates, remain insufficient \cite{keuning}. This raises the question of whether auto-grader feedback is in fact useful to students, or even utilized at all. In absence of such studies, instructors run the risk of dedicating considerable resources to developing tools that may not necessarily yield better learning outcomes. Such studies are especially timely when increasingly many educators explore ways to replace or augment human instructors' feedback with feedback generated by large language models (LLMs) \cite{prather2023robots,prather2024beyond}.

In this paper, we evaluate auto-graders from a Python course we offered in partnership with five community colleges in the United States during the 2022-23 academic year, with 199 student participants. In this course, we logged and analyzed students' navigation history, submission history, estimated time spent, and scores. The course is hosted on a proprietary learning and research platform called Sail()\footnote{\url{https://sailplatform.org}}.

For this course, students write source code in local files, and submit them by running a submitter executable file. Students then log on to the course website and check their scores and feedback, which are usually available within seconds. The auto-graders are containers in a Kubernetes cluster, developed by our faculty and teaching assistants, and run against student submissions. In this paper, we investigate the following two research questions:
\begin{enumerate}
    \item Do students consistently check auto-grader feedback every time they submit? (RQ1)
    \item Is checking auto-grader feedback associated with better learning outcomes, as measured by students' scores? (RQ2)
\end{enumerate}

\noindent We show that checking auto-grader feedback more frequently is associated with getting higher scores for programming assignments, and that checking auto-grader feedback between two consecutive submissions is associated with a higher probability of getting an improved score in the latter submission. We could not show that checking auto-grader feedback is associated with increased efficiency (as measured by estimated time spent). Further work is needed to examine our feedback template, experiment with varying templates, and better understand why some feedback templates might be more useful than others for learning outcome improvements.

This study is focused on auto-grader feedback's impact on adult students learning to write correct code, and excludes other learning objectives such as style and algorithmic performance. In the rest of this paper, all feedback is assumed to be provided by an auto-grader.

\section{Related Work}
This paper extends our prior work on student outcomes and working habits when engaged with auto-graded project-based CS/IT courses delivered through the Sail() Platform. Previously, we analyzed students' persistence, their course grades, and self-efficacy in an introductory programming classes, focusing on the delivery modality (i.e., online, in-person, cohort, synchronous, asynchronous) \cite{bogart2024factors,savelka2025ai}. We also explored different student approaches to working on programming projects (working habits), showing if and how certain habits could lead to better course performance \cite{an2021working,goldstein2019understanding}. This paper follows up on those work by investigating how students interact with auto-grader feedback, provided in a project-based introductory programming course.

Multiple literature review papers found that there is insufficient empirical evidence of auto-grader feedback's effectiveness in improving learning outcomes in programming education. One such systematic literature review of automated feedback generation tools developed by the year of 2015 \cite{keuning} shows a quarter of them had anecdotal or no evaluation at all; even the tools that had empirical evaluation (which was about a third of them) differed greatly in how they were evaluated, and often lacked detail on the methods and results. Another systematic literature review of such tools published between the years of 2017 and 2021 \cite{messer} shows most of these tools had been evaluated using surveys or by being compared to human graders rather than learning outcomes such as assignment grades, course grades, and third-party assessment grades \cite{pettit2015}.

Demonstrating auto-grader feedback's effectiveness for computer science education, and identifying which methods of feedback generation and presentation work better than others, can have far-reaching impacts. During the 2021-22 academic year, 108,049 Bachelor's degrees were conferred in Computer and Information Sciences and Support Services in the United States alone \cite{nsc2024}. Those numbers do not include students in other parts of the world, or students in other fields of study taking programming courses. At least 121 research papers on automated grading and feedback tools for programming education were published between the years of 2017 and 2021 \cite{messer}. With the growing number of students and institutions utilizing auto-graders, often in isolation \cite{pettit}, the importance of a shared understanding on how to generate and present feedback that is as effective as it is efficient cannot be overstated.

Generating effective auto-grader feedback for introductory courses is arguably more difficult than it is for intermediate or advanced courses, with failure rates of novice programming courses often exceeding 30\% \cite{sim2018,bennedsen2007}. This is because novice learners may even struggle to write code that compiles, not to mention parsing feedback or error messages provided by the auto-graders and by their console. Failure to understand console messages may cause failure to compile and submit code, which may in turn cause failure to receive any feedback at all. This adds all the more necessity to demonstrating that the feedback indeed helps students learn, so that fewer students feel left behind. At the time of our study, our auto-graders suffered from the same limitation, requiring successful compilation of the students' code before they can submit. We have since improved both our auto-graders and our learning platform such that the compilation requirement is removed for the initial programming assignments \cite{nguyen2024examining}, the effects of which we will extensively report on in the future.

A few studies show a positive impact of auto-grader feedback on student learning, but with limitations. Some had a small number of students working in groups and surveyed them after a short time period of observation \cite{kurniawan,kumar2005}, which may not be representative of how auto-graders are deployed in semester-long, potentially online, large courses. Some did not observe improvements in the treatment group's post-tests or post-assignments \cite{mitra}. Some did observe improvements after introducing auto-graders that provides immediate feedback \cite{gabbay,wang2011}, but it is unclear if the students indeed checked the feedback that was provided.

This study builds on those earlier findings by addressing one of their key limitations: being able to differentiate between students checking feedback and students not checking feedback, while providing feedback to everyone. We address that limitation by hosting the feedback on submission-specific web pages, and logging student navigation to those web pages, which allows us to evaluate our auto-grader feedback based on knowledge of which students presumably consumed them.

\section{Background}

\subsection{The Python Course}

Our Python course is entirely online and has 9 units, of which 5 are used by all the colleges included in this study:
\begin{enumerate}
    \item In \textbf{Hello World}, students practice how to make a submission and how to navigate to the feedback for that submission, using the Sail() Platform.
    \item In \textbf{Types, Variables, and Functions}, students practice creating the basic building blocks of Python, visualized using a simple calculator app with a graphical user interface.
    \item In \textbf{Iteration, Conditionals, Strings, and Basic I/O}, students implement a series of functions which, together with the starter code, become a report generating app.
    \item In \textbf{Data Structures}, students build on that report-generating app by manipulating lists, dictionaries, sets, and tuples.
    \item In \textbf{Object-Oriented Programming}, students practice OOP basics by declaring, instantiating, and extending simple classes.
\end{enumerate}

\noindent The course also has optional units in Software Development, Data Manipulation, Web Scraping, and Data Analysis. We excluded data from those units in this study because not every college used them.

Each unit has four types of modules: concepts, primers, quizzes, and projects. For this paper, we focused solely on projects as they employ auto-graders. There is exactly one project per unit, following a project-based education model \cite{kokotsaki}. Our prior work, \cite{savelka2023thrilled}, provides additional details on the course itself.

\begin{figure}[h!]
    \centering
    \includegraphics[width=1\linewidth]{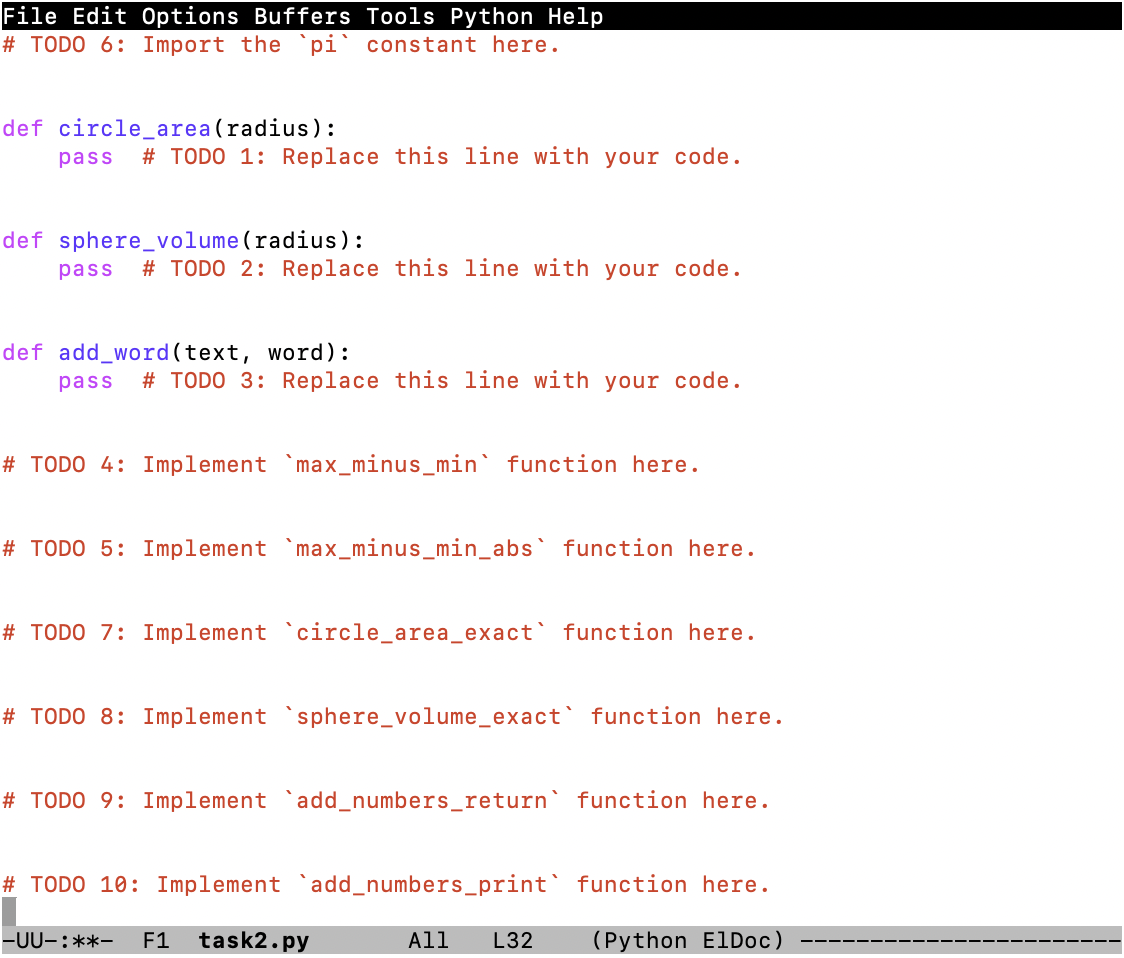}
    \caption{Example of Handout Code}
    \label{fig:handout}
\end{figure}

Projects are where students spend most of their time. For each project, students are presented with a real-world scenario, and handout code that scaffolds the scenario (See Figure \ref{fig:handout}). To complete the tasks within each project, which are usually spread across multiple Python files, students extend the code in each file (sometimes creating new files) drawing upon what they learned earlier in the unit.

\subsection{The Auto-grader Feedback}

\begin{figure*}[h!]
    \centering
    \includegraphics[width=1\linewidth]{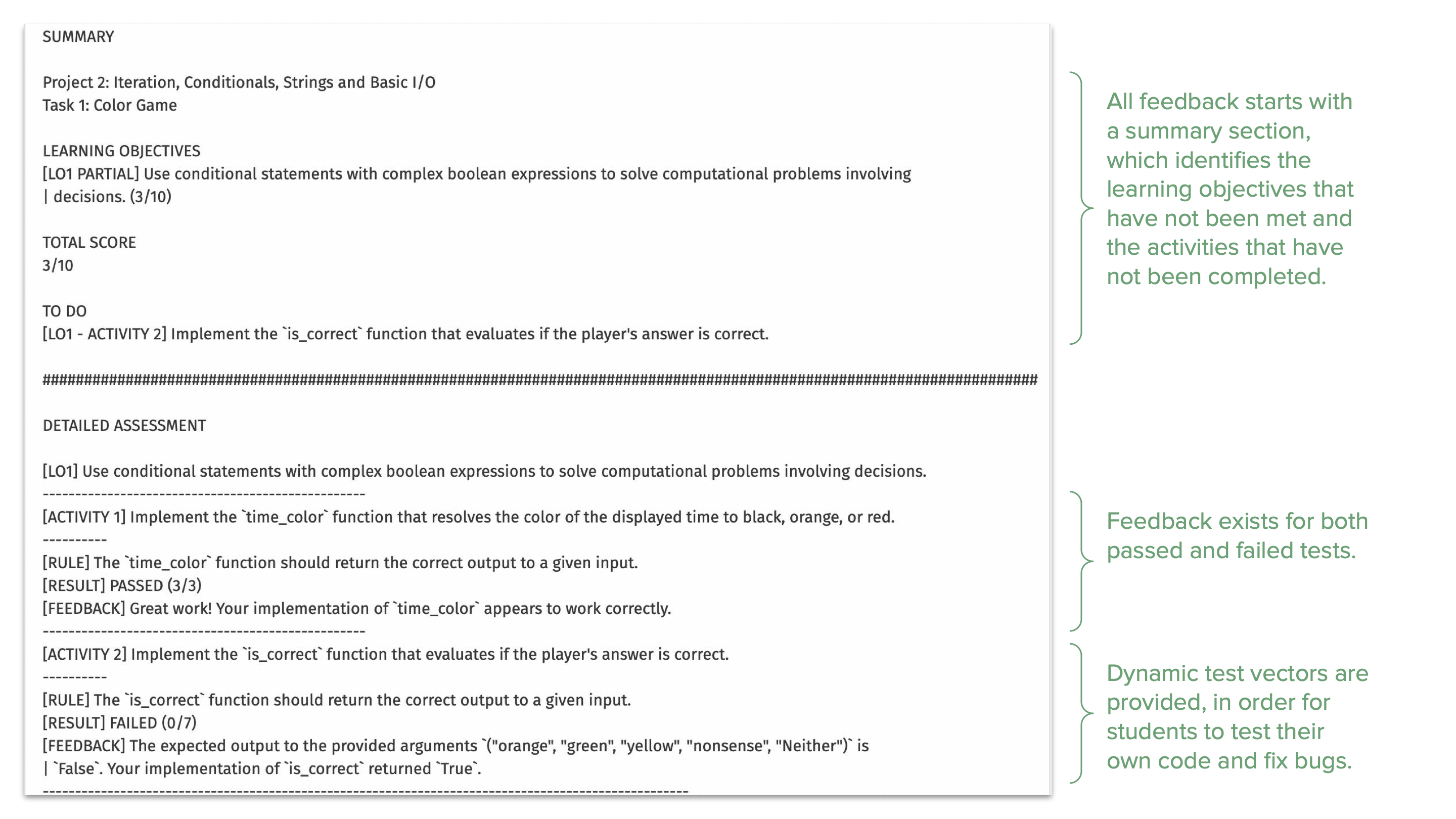}
    \caption{Example of Auto-grader Feedback for A Student's Submission}
    \label{fig:example_feedback}
\end{figure*}

Auto-grader feedback is generated within seconds for each submission. The feedback is hosted on a dedicated webpage accessible via a hyperlink on the submissions table for the task on the course website. The feedback is not provided anywhere else. The feedback is available only to the student who made the submission. The feedback is exclusively textual; an example is shown in Figure \ref{fig:example_feedback}.

\subsection{The Offerings}

When instructors offer the Python course, they may choose to add additional materials or activities beyond the scope of the original course. We do not collect data on those materials and activities.

Six instructors from five colleges in four American states offered the course in our one-year study. All instructors completed the course before teaching.

Two colleges offered the course asynchronously, which means all projects were open from the start to the end of the semester which spanned four months. The other three offered it synchronously, which means all students from the same section worked on the same project in a week, before they moved on to the next project next week.

\begin{figure}[h!]
    \centering
    \includegraphics[width=1\linewidth]{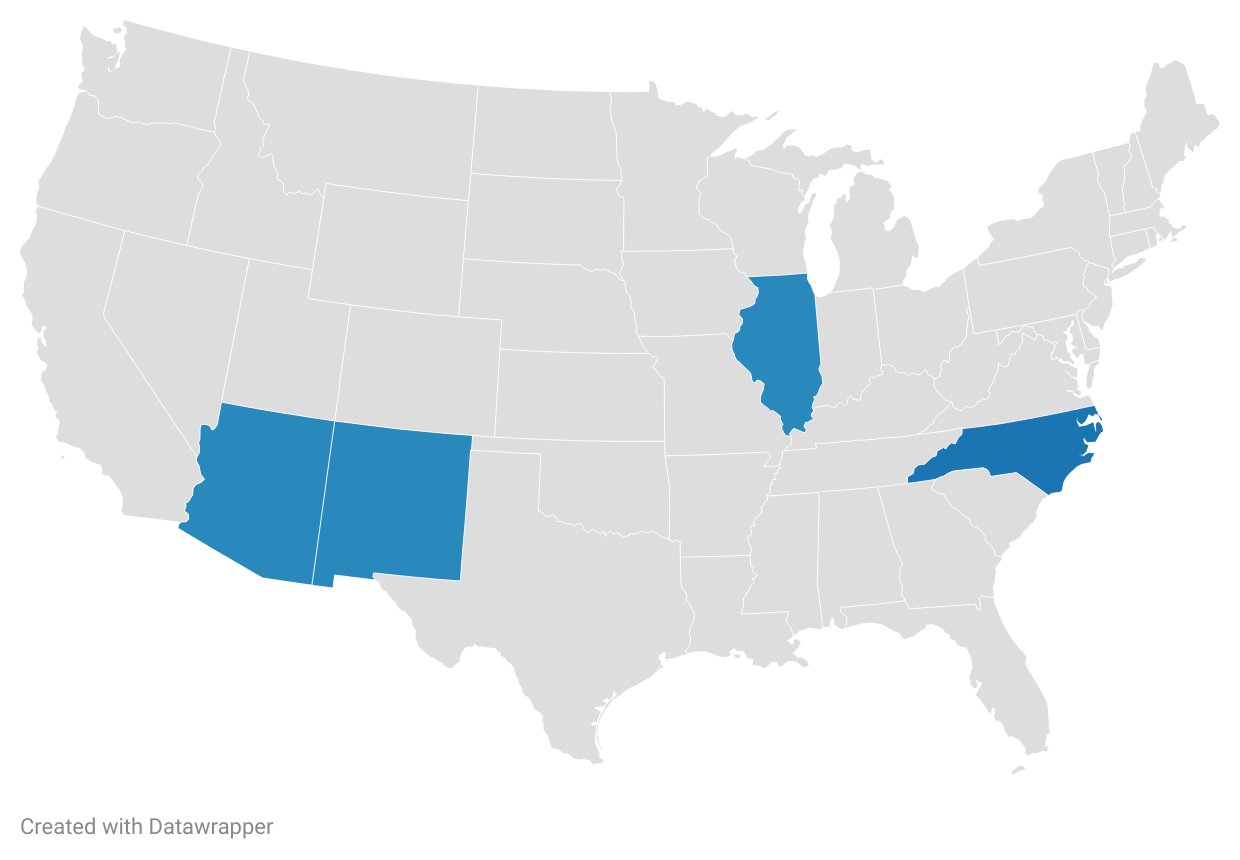}
    \caption{American States Represented by Participants;\newline
    Arizona (1), Illinois (1), New Mexico (1), and North Carolina (2)}
    \label{fig:map}
\end{figure}

\section{Dataset}
The dataset consists of records of 15 sections of the Python course, offered in Fall 2022, Spring 2023, and Summer 2023, at five community colleges in the United States, for 199 student participants in total. Records of students who dropped the course (at any time) or chose not to participate in the research are excluded from the study. The sections had between 3 and 28 students; most had around 15.

We analyzed data from the first five projects (which are used by all colleges in this study), and excluded data from later projects. The projects are zero-indexed, and will be referred to as \texttt{project0} through \texttt{project4} in later figures. Each project consists of 2--6 tasks, which could be distributed across multiple Python files. Students submit their code for each task, and get feedback for each submission. The number of submissions is not limited as long as the deadline has not passed.

We estimate how long a student spends on each project by counting distinct clock hours in which they navigate within that project's web pages or make submissions to that project's tasks. This is called \texttt{ProjectHours} in later diagrams.

We note the final score a student receives for each project when that project is due. This is called \texttt{ProjectScore} in later diagrams.

The colleges are anonymized as A, B, C, D, E.

We do not know if a student actually read the feedback for a submission, or how carefully they read it. We say that a student has checked the feedback if the event of them navigating to the webpage that hosts the feedback is logged by our web service.

\section{Results}

\subsection{Do students consistently check auto-grader feedback every time they submit? (RQ1)}

\begin{figure}[H]
    \centering
    \includegraphics[width=1\linewidth]{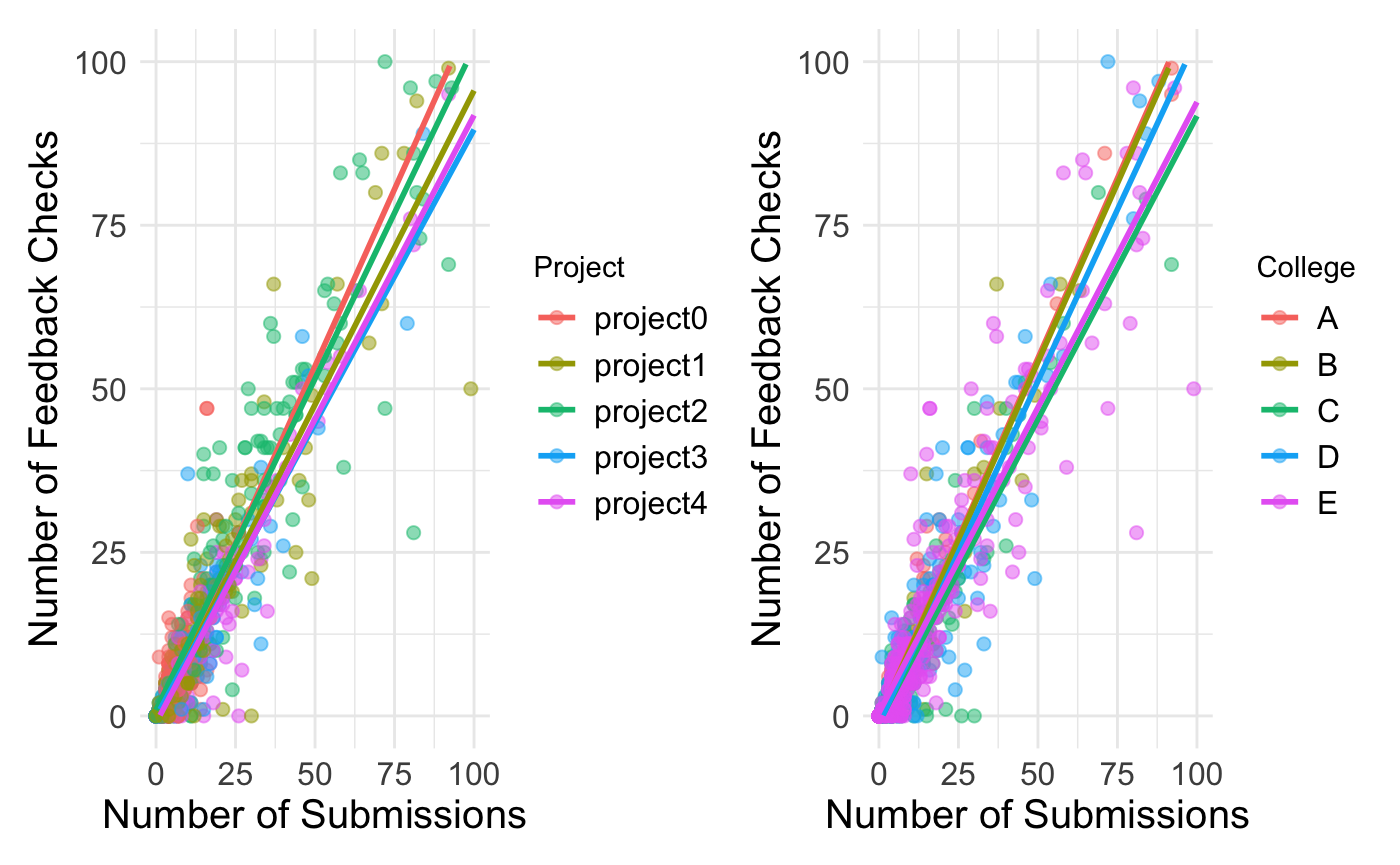}
    \caption{NFeedbackChecks v. NSubmissions;\linebreak
    Grouped by Project and by College}
    \label{fig:scatter_plot}
\end{figure}

Students check auto-grader feedback almost as often as they submit. Throughout the course, across five projects, students made an average of 66 submissions ($\sigma = 72$) and checked feedback an average of 64 times ($\sigma = 73$). The pattern also holds for each individual project and for each college, as shown in Figure \ref{fig:scatter_plot}, where the regression line for all subsets of the data (based on either project or college, as signified by the color) closely resemble each other. Their correlation is explored in the next subsection.

However, not all feedback is checked. Recall that feedback is generated by the auto-grader for each submission, on a webpage uniquely identified by the submission, and accessible only to the student who made the submission. 28\% of the feedback pages were never checked. 56\% were checked once. 16\% were checked more than once.

\begin{figure}[H]
    \centering
    \includegraphics[width=1\linewidth]{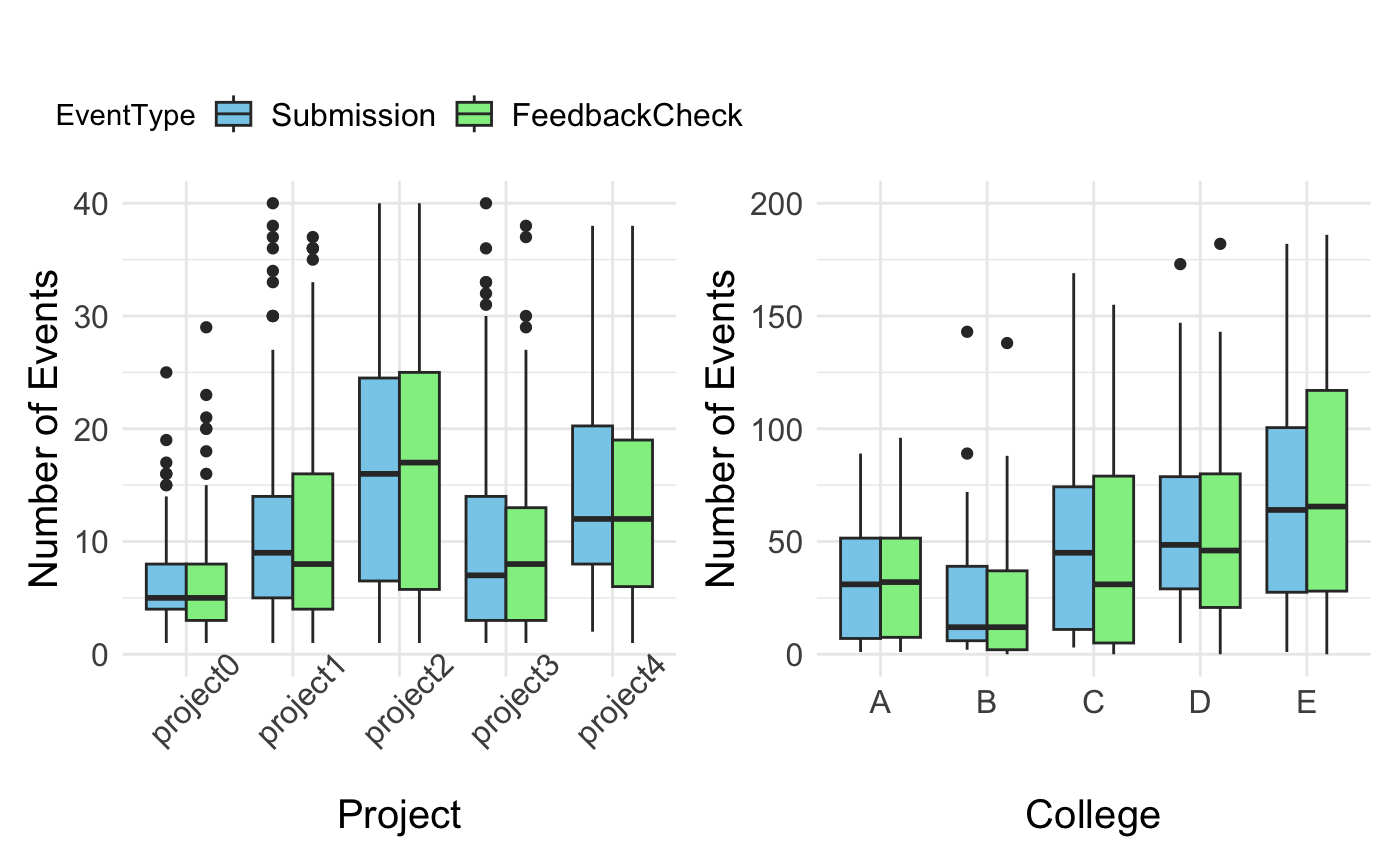}
    \caption{Number of Submissions and Feedback Checks per Student, Grouped by Project and by College}
    \label{fig:nevents_per_student}
\end{figure}

The number of feedback checks may be related to the assignment's difficulty. Students are about 3 times as active in \texttt{project2} as they are in \texttt{project0}, as shown in Figure \ref{fig:nevents_per_student}. Instructors and authors of the course expressed in interviews that \texttt{project2} is the most difficult of the projects. It also has one of the highest number of tasks (4), while \texttt{project0} and \texttt{project3} are two of the easiest projects with the lowest number of tasks (2). Students are second-most active in \texttt{project4}, which has the highest number of tasks (6).

\begin{figure}[h]
    \centering
    \includegraphics[width=1\linewidth]{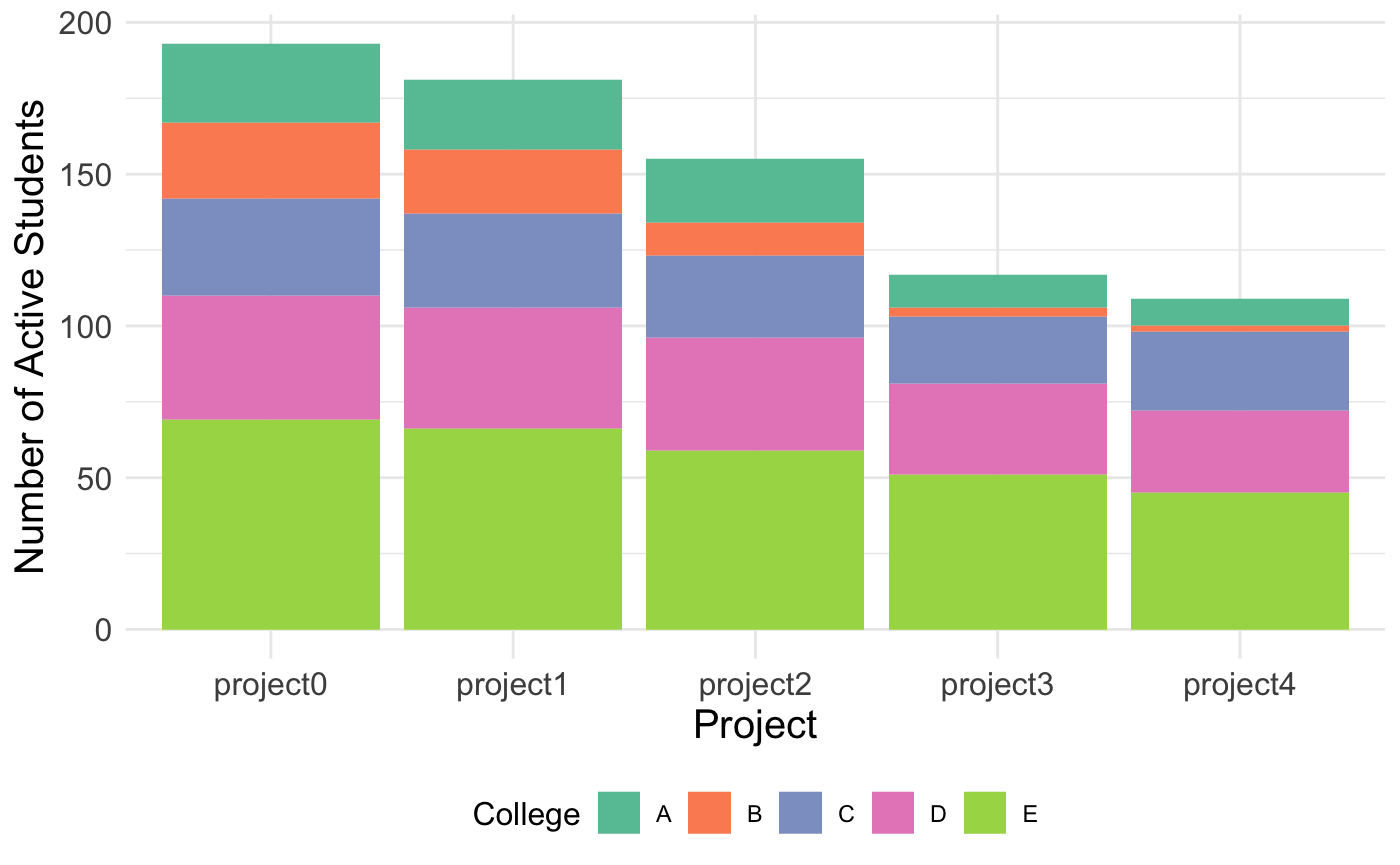}
    \caption{Number of Students Active for Each Project; \protect\linebreak making at least one submission to a project qualifies them as being active for that project}
    \label{fig:nstudents_by_project}
\end{figure}

Figure \ref{fig:nevents_per_student} also shows that students at College E are twice as active as students at College A, and even more so than students at College B. To better understand the variations (of the numbers of submissions and feedback checks) among the colleges, we analyzed how many students from each college attempted each project as their semesters progressed, as shown in Figure \ref{fig:nstudents_by_project}. We observed that almost all students from College B, and more than half of all students from College A, did not attempt \texttt{project3} or \texttt{project4}. Lower levels of participation would explain lower numbers of submissions and feedback checks.

\subsection{Is checking auto-grader feedback associated with better learning outcomes, as measured by students' scores? (RQ2)}

\begin{figure}[h]
    \centering
    \includegraphics[width=1\linewidth]{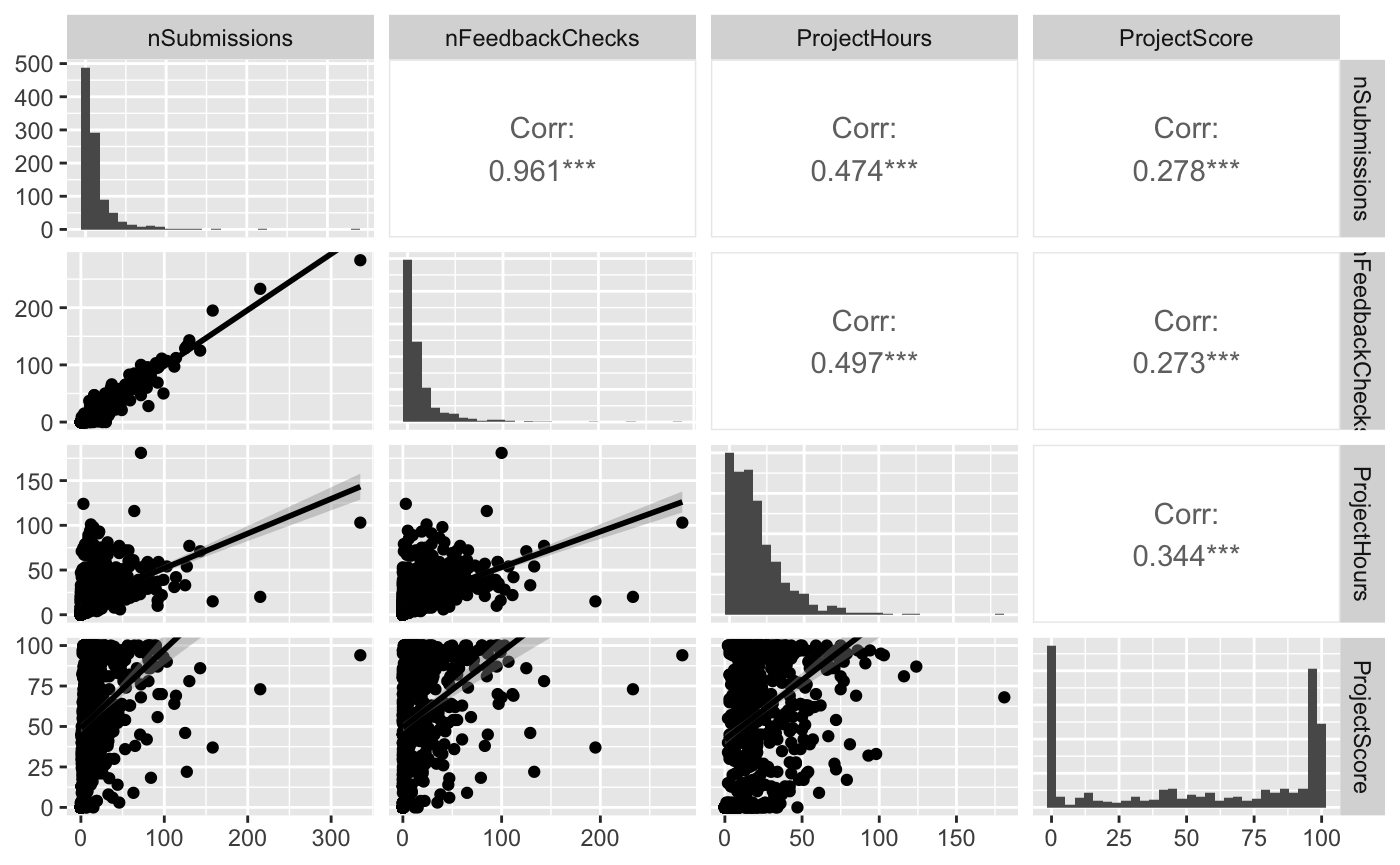}
    \caption{Correlation Matrix}
    \label{fig:correlations}
\end{figure}

We investigated the relationship between the number of submissions (\texttt{nSubmissions}) and feedback checks (\texttt{nFeedbackChecks}), the estimated time spent on a project (\texttt{ProjectHours}), and the project score (\texttt{ProjectScore}) for each student, using Pearson Correlations, as shown in Figure \ref{fig:correlations}.

We observed that \texttt{nSubmissions} and \texttt{nFeedbackChecks} have a strong and positive correlation ($cor = 0.961, p < 0.001$), which aligns with our earlier observation that students check feedback almost as often as they submit regardless of project and college.

\texttt{nSubmissions} and \texttt{ProjectHours} also have a strong and positive correlation ($cor = 0.474, p < 0.001$). We hypothesize that this is because it takes more time to make more submissions.

The strong and positive correlation between \texttt{nFeedbackChecks} and \texttt{ProjectHours} ($cor = 0.497, p < 0.001$) is a natural extension of the previous two correlations, but we could not determine if past a certain threshold students who check feedback more often per submission end up spending less time, which would support the hypothesis that checking feedback allows students to learn more efficiently.

\texttt{nFeedbackChecks} and \texttt{ProjectScore} have a moderately positive correlation ($cor = 0.273, p < 0.001$), which aligns with our hypothesis that feedback checks have a positive impact on scores, but the impact may instead come from spending more time on the project, as \texttt{ProjectHours} and \texttt{ProjectScores} also has a positive correlation that is a little stronger ($cor = 0.344, p < 0.001$).

To better understand the impact of checking feedback, we analyzed consecutive submissions made by the same student where the first submission did not get a full score, and observed if the student checked the feedback for the former submission, and if the latter submission got a higher score. We consider two submissions to be consecutive even if the student attempted another task or another project between the two submissions. Recall that there are multiple tasks in a project, and submissions are made to tasks.

Specifically, we call a submission \emph{non-maximal} if it does not receive the full score, and \emph{non-terminal} if the student makes another submission to the same task at any time in the future. We analyzed the set of all such non-maximal, non-terminal submissions, and computed the following conditional probabilities.


$P(Higher|Check) = 38.46\%$

$P(Higher|Not Check) = 33.77\%$

Students who check the auto-grader feedback after a non-maximal, non-terminal submission are more likely to score higher in the subsequent submission ($p=0.0063$ following a Fisher's Exact Test). This provides statistically significant evidence that checking auto-grader feedback likely has a positive impact on submission scores. Moreover, our result resembles that of a previous study \cite{gabbay} where the authors noted in 36\% of resubmissions for an online programming course with auto-grader feedback, students corrected their mistakes.

\section{Discussion}

Our study shows that students check auto-grader feedback almost as often as they submit code. Our study also provides evidence of a positive impact of checking auto-grader feedback on assignment scores and submission scores. Specifically, checking auto-grader feedback more frequently for an assignment is associated with getting a higher final score for that assignment, and checking auto-grader feedback between consecutive submissions is associated with a higher probability of getting a higher score in the latter submission.

However, our study is limited in that it does not take into consideration variations in the quality of feedback provided by our auto-graders: some feedback may be more useful than others. Further work is needed to categorize our feedback template using a framework such as Narciss' \cite{narciss}, examine our feedback generation process, and better understand how to systematically generate more useful feedback.

Our learning platform also lacked the ability for students to give feedback on the feedback they get from the auto-grader (a feature we are developing now). A thumbs-up/thumbs-down button, a place for students to enter plain text feedback, and a bookmark mechanism allowing students to communicate exactly where they feel improvements are needed, would help course authors identify and focus their attention on feedback that have been labeled as less useful, and help researchers better understand why some feedback might be more useful than other.

Instructors should consider actively encouraging students to check auto-grader feedback after each submission. This is particularly important for students who may be new to programming or auto-graded assignments. Instructors should also consider monitoring students' feedback-checking behavior. Students who rarely check feedback may need additional support or guidance. Finally, instructors should consider incorporating discussions about how to effectively use auto-grader feedback into their courses. This could include demonstrating how to interpret different types of feedback, sharing strategies for debugging based on feedback, and explaining how feedback relates to learning objectives.

\section{Conclusions}
In this study, we explored if and how often 199 students from five community colleges in the United States checked auto-grader feedback as they took the same introductory Python programming course, to see if their feedback-checking behavior seems related to their scores. Our results clearly show the relationship between the scores and students checking the feedback. The more often a student checks auto-grader feedback for a programming assignment, the more likely they are to get a higher score for that assignment. Furthermore, checking the auto-grader feedback for a non-maximal, non-terminal submission is associated with a 4.69\% higher probability of getting a higher score in the subsequent submission for the same task, than not checking it. Our findings are based on logging student navigation to the web pages hosting the individualized feedback, though we do not know if the student indeed read the feedback or how carefully they read it.

\section{Future Work}

Further analysis and categorization of our feedback, using frameworks such as Narciss' \cite{narciss}, will allow us to better understand what types of feedback are more useful than others. Keuning et al. also called for such a comparison in their systematic literature review \cite{keuning}.

Additionally, both researchers and instructors could benefit from instrumenting the learning platform for students to provide feedback on the feedback they get, so that researchers can evaluate their usefulness and instructors can improve their courses. As we enter an era where programming education is ubiquitous for learners at all levels, and generative AI is starting to generate course content and contextualized feedback, it becomes all the more necessary that we continue to demonstrate the usefulness (and usability) of auto-grader feedback provided to learners, so as to ensure that the next generation of programmers is well-prepare to enter today's technology workforce.








\section*{\uppercase{Acknowledgment}}


This material is based upon work supported by the National Science Foundation under Grant No. 2111305.

Hosting of the educational platform, and Azure credits for some student learning activities on the cloud service provider, are sponsored by Microsoft.

Recruitment of the participating community colleges was accomplished in collaboration with the National Institute for Staff and Organizational Development (NISOD).

\bibliographystyle{apalike}

\bibliography{example}

\begin{thebibliography}{}

\bibitem[An et~al., 2021]{an2021working}
An, M., Zhang, H., Savelka, J., Zhu, S., Bogart, C., and Sakr, M. (2021).
\newblock Are working habits different between well-performing and at-risk students in online project-based courses?
\newblock In {\em Proceedings of the 26th ACM Conference on Innovation and Technology in Computer Science Education V. 1}, pages 324--330.

\bibitem[Bennedsen and Caspersen, 2007]{bennedsen2007}
Bennedsen, J. and Caspersen, M.~E. (2007).
\newblock Failure rates in introductory programming.
\newblock {\em SIGCSE Bull.}, 39(2):32–36.

\bibitem[Bogart et~al., 2024]{bogart2024factors}
Bogart, C., An, M., Keylor, E., Singh, P., Savelka, J., and Sakr, M. (2024).
\newblock What factors influence persistence in project-based programming courses at community colleges?
\newblock In {\em Proceedings of the 55th ACM Technical Symposium on Computer Science Education V. 1}, pages 116--122.

\bibitem[Gabbay and Cohen, 2022]{gabbay}
Gabbay, H. and Cohen, A. (2022).
\newblock Exploring the connections between the use of an automated feedback system and learning behavior in a mooc for programming.
\newblock In {\em Educating for a New Future: Making Sense of Technology-Enhanced Learning Adoption: 17th European Conference on Technology Enhanced Learning, EC-TEL 2022, Toulouse, France, September 12–16, 2022, Proceedings}, page 116–130, Berlin, Heidelberg. Springer-Verlag.

\bibitem[Goldstein et~al., 2019]{goldstein2019understanding}
Goldstein, S.~C., Zhang, H., Sakr, M., An, H., and Dashti, C. (2019).
\newblock Understanding how work habits influence student performance.
\newblock In {\em Proceedings of the 2019 ACM Conference on Innovation and Technology in Computer Science Education}, pages 154--160.

\bibitem[Keuning et~al., 2018]{keuning}
Keuning, H., Jeuring, J., and Heeren, B. (2018).
\newblock A systematic literature review of automated feedback generation for programming exercises.
\newblock {\em ACM Trans. Comput. Educ.}, 19(1).

\bibitem[Kokotsaki et~al., 2016]{kokotsaki}
Kokotsaki, D., Menzies, V., and Wiggins, A. (2016).
\newblock Project-based learning: A review of the literature.
\newblock {\em Improving Schools}, 19.

\bibitem[Kumar, 2005]{kumar2005}
Kumar, A.~N. (2005).
\newblock Generation of problems, answers, grade, and feedback---case study of a fully automated tutor.
\newblock {\em J. Educ. Resour. Comput.}, 5(3):3–es.

\bibitem[Kurniawan et~al., 2023]{kurniawan}
Kurniawan, O., Poskitt, C.~M., Al~Hoque, I., Lee, N. T.~S., Jégourel, C., and Sockalingam, N. (2023).
\newblock How helpful do novice programmers find the feedback of an automated repair tool?
\newblock In {\em 2023 IEEE International Conference on Teaching, Assessment and Learning for Engineering (TALE)}, pages 1--6.

\bibitem[Messer et~al., 2024]{messer}
Messer, M., Brown, N. C.~C., K\"{o}lling, M., and Shi, M. (2024).
\newblock Automated grading and feedback tools for programming education: A systematic review.
\newblock {\em ACM Trans. Comput. Educ.}, 24(1).

\bibitem[Mitra, 2023]{mitra}
Mitra, J. (2023).
\newblock Studying the impact of auto-graders giving immediate feedback in programming assignments.
\newblock In {\em Proceedings of the 54th ACM Technical Symposium on Computer Science Education V. 1}, SIGCSE 2023, page 388–394, New York, NY, USA. Association for Computing Machinery.

\bibitem[Narciss, 2008]{narciss}
Narciss, S. (2008).
\newblock Feedback strategies for interactive learning tasks.
\newblock In Spector, J., Merrill, M., van Merrienboer, J., and Driscoll, M., editors, {\em Handbook of Research on Educational Communications and Technology}, chapter~11, pages 125--144. Lawrence Erlbaum Associates, Mahwah, NJ, 3rd edition.

\bibitem[{National Student Clearninghouse}, 2024]{nsc2024}
{National Student Clearninghouse} (2024).
\newblock Undergraduate degree earners: Academic year 2022-23.
\newblock Technical report, National Student Clearinghouse Research Center.

\bibitem[Nguyen et~al., 2024]{nguyen2024examining}
Nguyen, H.~A., Bogart, C., {\v{S}}avelka, J., Zhang, A., and Sakr, M. (2024).
\newblock Examining the trade-offs between simplified and realistic coding environments in an introductory python programming class.
\newblock In {\em European Conference on Technology Enhanced Learning}, pages 315--329. Springer.

\bibitem[Pettit et~al., 2015]{pettit2015}
Pettit, R., Homer, J., Holcomb, K., Simone, N., and Mengel, S. (2015).
\newblock Are automated assessment tools helpful in programming courses?
\newblock {\em ASEE Annual Conference and Exposition, Conference Proceedings}, 122.

\bibitem[Pettit and Prather, 2017]{pettit}
Pettit, R. and Prather, J. (2017).
\newblock Automated assessment tools: too many cooks, not enough collaboration.
\newblock {\em J. Comput. Sci. Coll.}, 32(4):113–121.

\bibitem[Prather et~al., 2023]{prather2023robots}
Prather, J., Denny, P., Leinonen, J., Becker, B.~A., Albluwi, I., Craig, M., Keuning, H., Kiesler, N., Kohn, T., Luxton-Reilly, A., et~al. (2023).
\newblock The robots are here: Navigating the generative ai revolution in computing education.
\newblock In {\em Proceedings of the 2023 Working Group Reports on Innovation and Technology in Computer Science Education}, pages 108--159. ACM.

\bibitem[Prather et~al., 2024]{prather2024beyond}
Prather, J., Leinonen, J., Kiesler, N., Benario, J.~G., Lau, S., MacNeil, S., Norouzi, N., Opel, S., Pettit, V., Porter, L., et~al. (2024).
\newblock Beyond the hype: A comprehensive review of current trends in generative ai research, teaching practices, and tools.
\newblock {\em arXiv preprint arXiv:2412.14732}.

\bibitem[Savelka et~al., 2023]{savelka2023thrilled}
Savelka, J., Agarwal, A., An, M., Bogart, C., and Sakr, M. (2023).
\newblock Thrilled by your progress! large language models (gpt-4) no longer struggle to pass assessments in higher education programming courses.
\newblock In {\em Proceedings of the 2023 ACM Conference on International Computing Education Research-Volume 1}, pages 78--92.

\bibitem[Savelka et~al., 2025]{savelka2025ai}
Savelka, J., Kultur, C., Agarwal, A., Bogart, C., Burte, H., Zhang, A., and Sakr, M. (2025).
\newblock Ai technicians: Developing rapid occupational training methods for a competitive ai workforce.
\newblock In {\em Proceedings of the 56th ACM Technical Symposium on Computer Science Education V. 1}.

\bibitem[Sim and Lau, 2018]{sim2018}
Sim, T.~Y. and Lau, S.~L. (2018).
\newblock Online tools to support novice programming: A systematic review.
\newblock In {\em 2018 IEEE Conference on e-Learning, e-Management and e-Services (IC3e)}, pages 91--96.

\bibitem[Wang et~al., 2011]{wang2011}
Wang, T., Su, X., Ma, P., Wang, Y., and Wang, K. (2011).
\newblock Ability-training-oriented automated assessment in introductory programming course.
\newblock {\em Comput. Educ.}, 56(1):220–226.

\end{thebibliography}



\end{document}